\documentclass[aps,prd,twocolumn,groupedaddress,showpacs]{revtex4}
\usepackage{graphicx}

\begin{document}
\title{Comment on ``Insensitivity of Hawking radiation to an invariant 
Planck-scale cutoff''}

\author{Adam D. Helfer}
\email{helfera@missouri.edu}
\affiliation{Department of Mathematics, University of Missouri,
Columbia, MO 65211}

\begin{abstract}  
I point out that the cutoff introduced by  
Agull\'o {\em et. al.}~\cite{ANGOP} 
has little impact on the trans-Planckian problem
as it is usually understood; 
it excludes only a 
small fraction of the problematic modes.
\end{abstract}

\maketitle

Investigation of Hawking's~\cite{Hawking1975} suggestion that black
holes radiate thermally remains one of the most active areas of 
research
in general relativity.  Yet despite an enormous amount of work, two
foundational questions remain:  the trans-Planckian problem (that is,
the dependence of Hawking's computation on modes of arbitrarily high
frequency); and the question of whether the neglect of possible 
quantum
gravitational effects is really correct.  (For a review, see
Ref.~\cite{ADH2003}.) Indeed, these concerns in part motivate the
ingenious attempts to find alternative derivations of Hawking's 
results
(e.g.~\cite{RobWil}).

In a recent interesting paper, Agull\'o {\em et al.}~\cite{ANGOP}
argued that Hawking's computation was in fact insensitive to a certain
Lorentz-invariant cutoff, drew parallels with the Unruh effect, and
suggested that this might overcome the trans-Planckian problem.  I
show here that this last hope is not fulfilled:  the cutoff removes 
only a negligible fraction of the relevant trans-Planckian modes.

Much of Agull\'o {\em et al.}'s treatment turns on concern over
the sense in which the trans-Planckian problem is invariant.  In the
absence of a well-known compact invariant characterization of this
problem, the authors in fact suggest using their manifestly invariant
cut-off as the {\em definition} of the relevant trans-Planckian
threshold; were we to accept this, implementing the cut-off would
resolve the difficulty.
Should we accept this definition?  That is,
does the cut-off really remove this problematic
element of the derivation of Hawking radiation?  

To avoid confusion, I will keep the conventional usage
of the phrase ``trans-Planckian problem,'' but I will
consider its invariance, as well as the central issue of
how much the cut-off helps with this troubling aspect of Hawking's derivation. 

We shall see that the conventional trans-Planckian problem {\em is}
invariant --- although a full understanding of this invariance requires
non-local considerations.  This invariant point of view will also make it
clear that only a substantial alteration of Hawking's original
argument could remove the problem; in particular, Agull\'o {\em et
al.}'s intentionally modest modification cannot resolve it. 
While this conclusion is negative, we
shall
find that examining the details of Agull\'o {\em et al.}'s argument
brings out a great deal of interesting physics in the way these general and
non-local considerations manifest themselves for specific
particle detections at particular places.

Following Agull\'o {\em et al.}, let us recall the main elements of
Hawking's idea.  We consider a linear massless scalar field on a
spherically symmetric space--time representing collapse to a black
hole of mass $M$.  We resolve the field by spherical harmonics, and
consider the propagation of the reduced radial modes.  (In fact, it is
known that the primary contribution to the Hawking effect is from the
s-wave sector, so one could restrict to that.)

The key is to consider what happens to a mode on its passage from the
distant past, through the region representing the incipient black 
hole,
and out to the future.  A mode whose angular frequency is $\omega '$ 
in
the past will give rise to a family of modes in the future, with the
Bogoliubov coefficients $\alpha _{\omega\omega '}$ and $\beta
_{\omega\omega'}$ representing the fractional contributions to 
outgoing
modes of angular frequencies $\pm\omega$, respectively.  Production of
quanta is governed by $\beta _{\omega \omega '}$, and so it is the
computation of these Bogoliubov coefficients which is the central
element in the analysis.  (The angular indices $l$, $m$ have been
omitted.)

Hawking showed that the significant contributions to $\beta _{\omega
\omega '}$ arise for $\omega\sim T_{\rm H}= (8\pi M)^{-1}$ (the 
Hawking temperature in natural units).
However --- and this is the 
trans-Planckian 
problem --- the values of $\omega '$ contributing mainly to these, 
that is, the frequencies of the modes which in the distant past are 
the precursors to the Hawking quanta, go like
\begin{equation}\label{tP}
 \omega'\sim Ce^{+u/(4M)}\, ,
\end{equation}
where $u$ is the retarded time.  That is, modes of a fixed frequency $
\sim T_{\rm H}$ in the distant future arise from modes of 
exponentially increasing frequencies in the distant past.  (On the 
other hand, these modes have, in the distant past, 
zero occupation numbers.  Thus it is vacuum fluctuations of 
exponentially high frequencies in the distant past which are supposed 
to be converted, by the Hawking process, into real quanta of moderate 
frequency in the future.)

I have written, as is conventional, of the modes' frequencies; of
course, measures of frequency presuppose a reference frame, for
frequency is not invariant.  The frequencies here are all measured 
with
respect to the asymptotic reference frame aligned with the incipient
black hole's energy--momentum.  (One could equivalently use the frame 
of
a fixed observer falling freely across the horizon --- only a finite
relative boost is involved in comparing these frames.) 
As noted above,
much of Agull\'o {\em et al.}'s motivation appears to derive from the
fact that the numerical frequencies of the precursors have, in
themselves, no absolute significance.
This, while true, gives the impression that the trans-Planckian 
problem
is a frame-dependent one and thus of dubious physical import.  
However,
the problem is an invariant one.

First, the matter in the distant past which will collapse to form the
hole defines an approximate reference frame.  Attempts to {\em
simultaneously} treat this matter and the Hawking effect by quantum
models will involve trans-Planckian wave-vectors whatever
frame one works in (and indeed this forms
an obstacle to developing a satisfactory theory of the back-reaction
of the Hawking effect on the space--time).  And second, because the
incipient hole is isolated, there is a well-defined way in which we 
can
compare the wave-vectors of the Hawking modes in the distant future
with those of their precursors in the distant past (one simply
parallel-transports along a path everywhere far from the matter).  
Thus,
while the numerical values of the frequencies are not invariant, the
{\em relative boost} of the precursors' wave-vectors to those of the
outgoing quanta is an invariant, and, in a frame in which the
outgoing quanta have moderate frequencies the precursors will be
trans-Planckian.  Another way of saying this is that there is an
invariant holonomy got by comparing parallel transport along two paths
from the distant past to the future, one the geometric-optics path 
taken
by the Hawking mode, and the other a path everywhere far from the
matter~\cite{ADH2001}.  See Fig.~1.

\begin{figure}
\includegraphics[width=.35\textwidth]{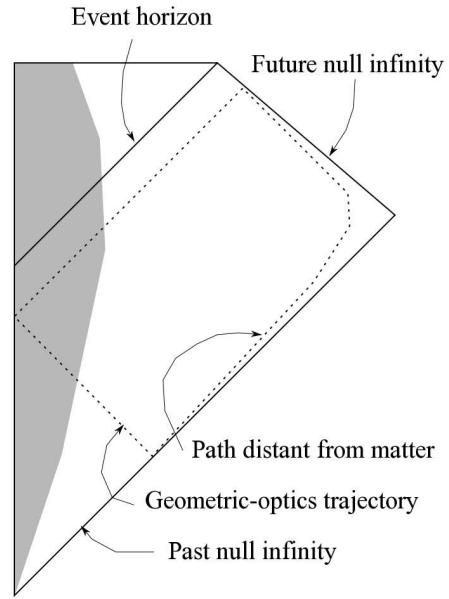}
\caption{Conformal diagram of the formation of a spherically symmetric
black hole.  The matter is shown shaded.  The trans-Planckian problem
is linked to a holonomy representing the difference in parallel
propagation along two paths, one
the geometric-optics trajectory of a Hawking mode and 
the other very distant from the matter.  The relative boost given
by this holonomy, which
carries moderate wave-vectors to trans-Planckian ones, is a physical
invariant.}
\end{figure}

While these trans-Planckian wave-vectors in the distant past
correspond to vacuum fluctuations, they can
give rise to real physical effects via standard quantum-theoretic
principles.  Thus real trans-Planckian physics can enter in the
consideration of quantum measurement issues involving correlations of
detectors in the reference frame in the past with Hawking quanta in 
the
future~\cite{ADH2004a}, or, when nonlinear couplings are considered, 
by
altering the detailed balance of the virtual creation and annihilation
processes --- some of the energy--momentum of these virtual processes
can be lost to the incipient hole, resulting in ``bubble'' diagrams
failing to close and the production of real ultra-energetic
quanta~\cite{ADH2004b}.  The physics in each case is invariant
because in each
case what matters is the comparison of the reference frame in the past
to that in the future via the two different routes giving rise to the
holonomy.

But the problem is still more severe, because the wave-vectors of the
problematic modes do not merely cross the Planckian threshold; they
diverge.  The trans-Planckian problem would be manifest {\em in any
fixed} reference frame, 
for a sequence of wave-vectors which diverges to
infinity in one reference frame will do so in any.  This applies 
whether
the problematic modes are examined in the distant past, or near the
horizon.  Thus while the particular frequencies considered are
frame-dependent, the trans-Planckian problem is an invariant one, even
if we do not wish to appeal to the existence of an
approximate physically preferred
frame in the distant past, or the non-local holonomy.

It is worth emphasizing that these arguments about propagation
do not establish any overt remarkable behavior of the field 
itself or the two-point function either near
the horizon or in the distant past.  What they do show is that {\em if}
one selects the field modes which will, in the future, develop into
Hawking quanta, {\em those} modes behave problematically in these earlier
regimes.  This is again the statement that the trans-Planckian problem
is non-local, since it depends on selecting modes in an early
regime on the basis of behavior at a later one.  We shall see below how
this contrasts with the characterization of trans-Planckian modes
proposed by Agull\'o {\em et al.}

All of these comments turn on something well-known and, I believe,
uncontroversial:  that Hawking's original computation of the modes'
behavior is mathematically correct, and that, according to it, the
Hawking quanta do arise from trans-Planckian precursors.
This means that
the trans-Planckian problem is very much at the heart of Hawking's 
original analysis.  If we believe Hawking's computation of the 
Bogoliubov coefficients, then the Hawking quanta do arise from 
trans-Planckian vacuum fluctuations in the distant past, and no 
alternative derivation can circumvent this; one would need to 
suppose different physics applies.  (See 
e.g.~\cite{Unruh1995,Jacobson1993,JacMatt,BMPS} for such
suggestions.)

Agull\'o {\em et al.} introduce a cutoff in an invariant way which
reproduces Hawking's results (with small corrections).  They suggest
their cutoff solves the trans-Planckian problem; however, it is
evident from the discussion in the previous paragraph 
that the very nearness of their
result to Hawking's means that there can be little change in the
contributions of those trans-Planckian modes giving rise to the
Hawking quanta.  I shall show here that what Agull\'o {\em et al.}'s
arguments actually establish is that for the Hawking quanta produced
in any interval of retarded time $ \Delta u$ a few Hawking-periods
long, there is a reference frame in a dimensionally reduced
space--time (not the physical space--time itself), relative to which
the precursors for the Hawking radiation in $\Delta u$ are not
trans-Planckian.  However, there is no single such frame (even in the
dimensionally reduced space); as later and later intervals $\Delta u$
are considered, the corresponding frames are boosted exponentially.

Agull\'o {\em et al.} argue as follows.  They assert that a detector 
held just outside the horizon of the incipient black hole will respond 
in the same way as an Unruh detector of the corresponding acceleration 
$a$ in Minkowski space.  In fact, as will be discussed below, while
this assertion is a common one, it is not fully justified.  But in
fact Agull\'o {\em et al.}'s argument does not really 
depend on such a direct physical correspondence, but rather on the 
mathematical correspondence between the two-point functions of the two 
systems.\footnote{In this connection, one should note that, had
they really depended on the physical correspondence, they would have 
had to account for the ultra-high accelerations of observers hovering 
just outside the horizon:  but Agull\'o {\em et al.} consider an Unruh 
detector with acceleration $a=(4M)^{-1}$.} This is given by formulas
(8) and (20) in their paper, which are standard.

The authors examine the 
expected number of Unruh quanta over a proper time interval $\Delta 
\tau$ with $a\Delta \tau\lesssim 1$.  They argue that in the local 
frame of the detector, the precursors of these Unruh quanta have only 
moderate frequencies, and that recasting the entire computation in 
terms of the two-point function $G(x_1,x_2)$ (from which $\beta 
_{\omega \omega '}$ can be extracted), once can impose the invariant 
cutoff
\begin{equation}\label{ek}
  |G(x_1,x_2)|<l_{\rm P}^{-2}
\end{equation}
(where $l_{\rm P}$ is the Planck length) without substantially 
affecting the result.  They then argue that Eq.~(\ref{ek}), carried 
over into the black-hole case, would give a cutoff theory in which 
Hawking quanta (slightly modified) were produced but with no 
trans-Planckian problem.

While Agull\'o {\em et al.} are certainly correct that Eq.~(\ref{ek})
provides an invariant trans-Planckian cut-off, it is important to
recognize that it has, on its face, little to do with the
trans-Planckian modes which are problematic for the Hawking process.
(Were we to try to formulate the usual trans-Planckian problem in terms
of two-point functions, we should expect to use at least two pairs of
widely separated points:  one pair in the distant past or at the
horizon, to detect the precursors' frequencies, and the second pair in the
distant future, to detect the Hawking quanta's frequencies.)  In order to
understand what the significance of restriction~(\ref{ek}) is, let us
examine the authors' subsequent analysis.

Leaving aside the black-hole case for the moment, Agull\'o {\em et 
al.}'s argument does indeed show that the 
precursor of an Unruh quantum is, in the frame in which the quantum is 
detected, of frequency $\sim a$.  On the other hand, if one wants to 
consider the response of the detector over an interval of proper time 
$\Delta\tau\gg a^{-1}$, then the boosts of the precursors, with 
respect to a 
fixed reference frame, over that interval vary by $\sim \exp
(a\Delta\tau )$.  
Thus while over any short time one does not have to invoke 
ultra-high-frequency modes to explain the Unruh effect, a treatment of 
it over long times does require exponentially large boosts.

Accepting now the correspondence with the black-hole case, we see that 
what Agull\'o {\em et al.} have shown is that over any short interval 
$\Delta u$ of retarded time, there is a frame in which Hawking quanta 
detected near the horizon have precursors of only moderate 
frequencies.  However, if $\Delta u_1$, $\Delta u_2$ are two such 
intervals separated by $\Delta u_{12}=\Delta u_1-\Delta u_2\gg 4M$,
then the precursors of the Hawking quanta in the
intervals are relatively boosted by a 
factor $\sim \exp \Delta u_{12}/(4M)$.  

One should remember, too, that the analyses of detectors set out at 
different angular positions around the hole will require boosts in 
different directions, with the boosts of antipodal detectors 
oppositely directed.  Therefore even over a fixed relatively short 
retarded time interval $\Delta u$, the boosts required to find 
cis-Planckian precursors for the Hawking quanta registering in 
antipodal detectors will differ from one another by an exponentially 
increasing factor, $\sim\exp 2u/(4M)$. 
It is only in the dimensionally reduced problem, where we
consider (say) the s-wave sector as a theory in a two-dimensional 
space, that this issue does not appear.

\begin{figure}
\includegraphics[width=.35\textwidth]{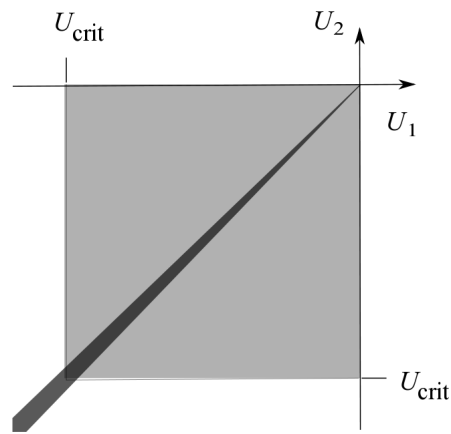}
\caption{The part of the $U_1$--$U_2$ plane contributing to the
trans-Planckian problem and the sector excluded by the invariant
cutoff.  The trans-Planckian region is the square $U_{\rm
crit}<U_1,U_2<0$; only a narrow sector
(shown darkened and with its width exaggerated), of angle $\sim m_{\rm 
P}/
(32\pi M)$, is
excluded by the condition of Agull\'o {\em et al}.  
By contrast, virtually all of the square is required for the
modeling of the detection of Hawking quanta.}
\end{figure}

We can see explicitly that Agull\' o {\em et al.}'s cutoff excludes 
only
a small fraction of the relevant trans-Planckian modes by examining 
the
form they actually use in the black-hole case.  This is cast in 
terms of the Kruskal retarded time
\begin{equation}\label{kt}
U=-(4M)\exp -(u/4M )\, ,
\end{equation} 
which is a smooth null coordinate in the vicinity of the
horizon.\footnote{Strictly speaking, the Kruskal coordinate itself is
defined only exterior to the matter, but there is a natural extension 
of
it inwards as a null coordinate respecting the spherical symmetry, and
this suffices for the analysis; cf. Ref.~\cite{ADH2003}.}
Note that $U<0$; the limit $U\to 0$ corresponds to approach to the 
event
horizon.

If we ask when the trans-Planckian problem sets in, we see from
relation~(\ref{tP}) that there will be a retarded time $u$, say $u_{\rm
crit}$, beyond which the precursors of the Hawking quanta have
trans-Planckian frequencies.  This will correspond,
according to Eq.~(\ref{kt}), to a value $U_{\rm
crit}<0$ of Kruskal retarded time; the Hawking quanta's precursors will
be trans-Planckian for $U_{\rm crit}<U<0$.

When we consider the detection of Hawking quanta in the distant future
in terms of two-point functions, these can be expressed either
using conventional retarded times $u_1$, $u_2$ or Kruskal
coordinates $U_1$, $U_2$.  A detector for quanta of frequency
$\omega$ will operate over a finite interval of retarded time $\Delta u
\gtrsim 1/\omega$;
the expected number of quanta it registers will be an integral of the
two-point function (weighted by a function encoding
the sampling profile of the detector) 
for $u_1$, $u_2$ in the interval.
Except for a restricted class of detectors operating
early on, this interval will be entirely within the regime $u>u_{\rm
crit}$ corresponding to trans-Planckian precursors.  In Kruskal
coordinates, for detectors with operating Kruskal times $U> U_{\rm
crit}$, the relevant arguments of the two-point function will be
entirely within the trans-Planckian region $U_{\rm crit}<U_1,U_2<0$ of
the $U_1$--$U_2$ plane.

The cutoff used by Agull\'o {\em et al.} in the black-hole case, 
their paper's inequality (11), can be written as
\begin{equation}\label{pk}
  (U_1-U_2)^2 >m_{\rm P}^2 (16\pi M)^{-2}U_1U_2
  \, , 
\end{equation}
where $m_{\rm P}$ is the Planck mass.
This cutoff
does excise a region from the $U_1$--$U_2$ 
plane, but it is a small one; a little algebra shows that (for $M\gg 
m_{\rm P}$) the condition is equivalent to
\begin{equation}
 |U_1/U_2 -1|> m_{\rm P}/(16\pi M)\, .
\end{equation}
Thus only a small fraction $\sim m_{\rm P}/(32\pi M)$
of the trans-Planckian regime in the $U_1$--$U_2$ plane is excluded.  
(See Fig. 2.)  By 
contrast, if a detector in the distant future responds over an 
interval $|u_2-u_1|\sim n(2\pi /T_{\rm H}) =n(16\pi ^2 M)$ 
(that is, $n$ times the period of a characteristic Hawking quantum), 
the 
corresponding ratio is $U_1/U_2\sim \exp \pm 4\pi ^2 n$ --- so nearly 
the whole square $U_{\rm crit}<U_1,U_2<0$ is required for modeling 
simply the detection of individual Hawking quanta
($n\sim 1$). 
Were we to consider 
correlations between quanta over extended intervals, or the detection 
of lower-frequency quanta, more of the square would be needed.
Thus, of the trans-Planckian regime which actually figures in the 
Hawking analysis, only a small portion is removed by
the condition of Agull\'o {\em et al.}

Finally, let me turn to the physical correspondence
between the Hawking and Unruh analyses.  It is often asserted that a
detector held just outside an incipient black hole must, by the
equivalence principle, respond like an Unruh detector.  
While there certainly is a connection, it is not at all such a simple 
correspondence.
A fast way of seeing this is to note that the Unruh analysis
is time-symmetric, but Hawking's analysis is definitely not
(it uses very strongly that the black hole is formed by collapse; a
static black hole would not Hawking--radiate).  What actually goes
wrong with the correspondence
argument is that the equivalence principle applies only
locally, but the scales required to compute the Hawking effect are
large enough to detect the difference of the quantum state from the
Minkowski vacuum.  (See Ref.~\cite{ADH2003} for a quantitative
treatment.)

\begin{acknowledgments}

I thank Professors Agull\'o, Navarro-Salas, Olmo and Parker for many
detailed and useful comments.  
\end{acknowledgments}

\bibliography{../references}

\end{document}